\documentclass[useAMS,usenatbib,asm]{mn2e}
\usepackage{graphicx,fleqn,times}

\usepackage{graphicx}
\usepackage{color}

\def\today{\ifcase\month\or
 January\or February\or March\or April\or May\or June\or
 July\or August\or September\or October\or November\or
 December\fi\space\number\day, \number\year}

\def\todmy{\number\day\space\ifcase\month\or
 January\or February\or March\or April\or May\or June\or
 July\or August\or September\or October\or November\or
 December\fi\space\number\year}

\newcommand{\tal}{\it et al. \rm}

\title{Unravelling the mystery of the M31 bar}
\author[E.~Athanassoula, Rachael Lynn Beaton]
       {E.~Athanassoula$^{1}$, Rachael Lynn Beaton$^{2}$
\\
$^1$ LAM, Observatoire Astronomique de Marseille Provence,
2 Place Le Verrier, F-13248 Marseille Cedex 4, France \\
$^{2}$ Department of Astronomy, University of Virginia,
Charlottesville, Virginia 22903-0818, USA\\
}

\date{Accepted .
      Received ;
      }

\pagerange{\pageref{firstpage}--\pageref{lastpage}}
\pubyear{2006}

\begin{document}

\maketitle

\label{firstpage}

\begin{abstract}
The inclination of M31 is too close to edge-on for a bar component to be
easily recognised and is not sufficiently edge-on for a boxy/peanut
bulge to protrude clearly out of the equatorial plane. Nevertheless,
a sufficient number of clues allow us to argue that this galaxy is
barred. We use fully self-consistent $N$-body simulations of barred
galaxies and compare them with both photometric and kinematic 
observational data for M31. In particular, we rely on the near
infrared photometry presented in a companion paper. We compare
isodensity contours to isophotal 
contours and the light profile along cuts parallel to the galaxy
major axis and offset towards the North, or the South, to mass
profiles along similar cuts on the model. All these comparisons,
as well as position velocity diagrams for the gaseous component, give
us strong arguments that M31 is barred. We compare four fiducial
$N$-body models to the data and thus set constraints on the parameters
of the M31 bar, as its strength, length and orientation. Our `best'
models, although not meant to be exact
models of M31, reproduce in a very satisfactory way the main relevant
observations. We present arguments that M31 has both a classical
and a boxy/peanut bulge. Its pseudo-ring-like structure at roughly $50'$ is
near the outer Lindblad resonance of the bar and could thus be an
outer ring, as often observed in barred galaxies. The shape of the
isophotes also argues that the vertically thin part of the 
M31 bar extends considerably further out than its boxy bulge, i.e. that
the boxy bulge is only {\it part} of the bar, thus confirming
predictions from orbital structure studies and from previous $N$-body
simulations. It seems very likely that the backbone of M31's boxy
bulge is families of periodic orbits, members of the x$_1$-tree and
bifurcating from the x$_1$ family at its higher order vertical
resonances, such as the x1v3 or x1v4 families.  
\end{abstract}

\section{Introduction}

It is very easy to recognise bars in disc galaxies which are far from
edge-on. This, however, is not true for
discs seen  edge-on, the safest approach being to use
kinematics \citep*{km95, mk99, bf99, ba99, ba05,
ab99, cb04}. Yet kinematics are
often not available, and thus one has to rely on morphology
and photometry alone. A large number of properties (see \cite{a05},
hereafter A05, for a global discussion and a compilation of 
references) argue that a boxy/peanut bulge is a part of a bar seen
edge-on. Thus, the mere presence of such a bulge
is telltale of the existence of a bar component (but see \cite{pags02},
arguing that other rotating bisymmetric perturbations could also
form boxes). If the galaxy is observed side-on
(i.e. edge-on with the line 
of sight along the bar minor axis), then the boxy or peanut feature is
clearly visible. If, however, the bar is seen end-on (i.e. edge-on
with the line of sight along the bar major axis), then the bar has the
outline of a classical bulge\footnote{Following A05, we distinguish
  throughout this paper three types of bulges. {\bf Classical 
  bulges} were formed in the early
  stages of the galaxy formation process, before the present discs,
  from gravitational collapse, or hierarchical merging of
  smaller objects and the corresponding dissipative processes. {\bf
    Boxy/peanut bulges} are part of the bar, i.e. they are constituted
  by stellar disc material, and they form naturally from the long-term
  evolution of the stellar component of bar unstable discs, as
  witnessed in a very large number of simulations. Finally, {\bf
    disc-like}, or {\bf discy bulges} form out of gas that 
  concentrates to the inner parts of the disc under the influence of the
  gravitational torque of a bar, and subsequently forms stars.} 
  and it can not be recognised. Luckily, even 
deviations as small as 10$^\circ$ from end-on will reveal the
bar/peanut feature \citep*[][A05]{ldp00b}. Evidence for bars in
edge-on systems can also be found from their photometry \citep{ldp00b}.
Indeed, in such cases, cuts along the equatorial plane
reveal plateaux on either side of the center. Such plateaux are also
seen in similar cuts in barred galaxy simulations (Fig. 6 in \cite{am02}, 
hereafter AM02) where it can be verified that they
extend roughly to the end of the bar. The existence of such plateaux on
equatorial cuts is, however, not unambiguous  evidence for the
existence of a bar component, since these features can also be due to lenses.   

Although considerable effort has been put into recognising bars in
edge-on systems, very little has been done for near-edge-on
systems. \cite{ba05} have described the position
velocity diagrams (hereafter PVDs)
of simulations viewed under such an orientation, as well as the
corresponding $V$, $\sigma$, $h_3$ and $h_4$ profiles (see their
Fig. 4). If, however, kinematical data are not available, then
recognising a bar in such orientations is more difficult than in
edge-on galaxies. Indeed, 
in such cases the boxy/peanut bulge does not clearly stick
out of the equatorial plane, so that this diagnostic can not
be used. Furthermore, no photometric analysis of simulations viewed
under such orientations is available to predict what one should
expect for the surface density profiles. 

Although many disc galaxies are viewed at angles near to edge-on and a
large fraction of them should have boxy/peanut bulges and bars, very few
have been discussed in the literature. NGC 7582 has an inclination of 
approximately $65^{\circ}$ and, when viewed in the near infrared
\citep{Quillen+}, allows for visual identification of both the bar and the 
peanut. Similar comments can be made for NGC 4442 \citep{BetGal},
viewed at approximately $72^{\circ}$. 
M31, our nearest big neighbour, is near to edge-on and exhibits
isophotal shapes similar to those of NGC 7582 and NGC 4442. 
Could it be barred? We will aim to answer this question by comparing
results from $N$-body simulations to the relevant data.

M31 (NGC 224) is an Sb type spiral galaxy, which has an 
inclination angle of about $77^{\circ}$. The isophotes of M31 have a twist
in the inner parts \citep{lindblad, hoken} and in the
outer parts \citep{WK}, the latter presumably due
to a warp. In the main region 
of the disc, however, the position angle is well defined and nearly
constant \citep[see Fig. 7 of][]{hoken}. We will thus
henceforth adopt a fiducial disc position angle of $38^{\circ}$ 
(\cite{deVauc} and later work). Further in, in the region often
referred to as the `bulge region', the position angle is $10^{\circ}$
to $20^{\circ}$ larger (\cite{Richt+Hogner} and later
work). The distance of M31 is estimated to be 783 kpc 
\citep{holland}, which makes $10'$ roughly equal to 2.28 kpc. In earlier
studies, however, it was assumed that $10'$ is roughly equivalent to 2
kpc. For this reason, and in order to avoid confusion, we will here
use arc minutes to measure distances. Due to its proximity and
its interesting structure, M31 has been the subject of a very large
number of studies, of which a complete list of references
is beyond the scope of this paper. Amongst the older photometry, 
we have used isophotal results from \cite{lindblad} and from \cite{deVauc}, 
as well as position angle measurements from \cite{hoken}.
More recently, M31 was observed by the 2MASS facility as part
of the ``6X Survey'', going 1 magnitude deeper than the original
2MASS survey\footnote{See the 6X description at \\
http://www.ipac.caltech.edu/2mass/releases/allsky/doc/}.  
These images, in $J$, $H$ and $K_s$ were analysed by
Beaton \tal (2006, hereafter Paper II) revealing important
morphological properties, especially of the bulge. It is these observational
 results that originally motivated the work described here. 

The idea that M31 is barred is not new. It was initially proposed by 
B. \cite{lindblad}, based on the analysis of the isophotal shapes of a
deep red exposure. Approximating the isophotes by ellipses, he found
that, on the plane of the sky, their major axis is at an angle of
roughly 10$^\circ$ south of the galaxy major axis and that the average
projected 
axial ratio is 1.6. He proposed that the bulge is in fact a
triaxial bar with axial ratios 1:0.56:0.28 or 1:0.56:0.36 and
semi-major axis length around $15'$.   
Yet, the existence of a bar in M31 was generally not accepted
and soon forgotten, so that hardly any of the very numerous
subsequent papers on M31 mention this possibility. 
\cite{Stark} returned to this conclusion with his calculation 
of a one-parameter family of
triaxial models for the M31 bulge. Their axial ratio in the
galaxy plane is about 1.6
to 1.8, depending on the angle of the bar major axis with the galaxy
major axis. The two solutions initially proposed by \citeauthor{lindblad} are
members of this family. To distinguish between \citeauthor{Stark}'s models, 
\cite{StarkBin} modeled the gas flow in the inner couple of
arc minutes using a Ferrers ellipsoid and a spherical component. They
propose that, measured on the plane of the galaxy, the angle between
the bar semi-major axis and the projection of the line of sight onto
the plane of M31 is about 20$^\circ$.

In this paper we will examine the possibility of M31 being a barred
galaxy based partly on the photometric results of 
paper II and partly on observed kinematics \citep{BrSh, BrBu,RF}. 
The wealth of information provided by these results will allow 
us to go further than just debate on the possible existence of a
bar. We will thus aim to find some of the main properties of this
bar. It should, however, be clear that we will not present a 
complete model of M31, which would be well beyond the scope of the
present paper, but we will instead try to investigate all the available
observational clues in order to characterize and constrain the bar
properties. This paper is organised as follows. 
Section~\ref{sec:simul} describes the simulations and 
Section~\ref{sec:M31} compares them to the photometrical observations
of Paper II. The kinematics are presented in
Section~\ref{sec:kinematics}. We discuss our results in
Section~\ref{sec:discuss} and briefly summarise in
Section~\ref{sec:summary}.   

\section{Simulations}
\label{sec:simul}

Both observed and simulated bars come in a large variety of lengths,
strengths and shapes. We will thus not restrain ourselves to one
single simulation to compare with the observations, but use four
distinct cases, 
hoping to be able to set constraints on the parameters of the bar
\footnote{All simulations described here have live haloes. This is
  necessary since, as shown by \cite{a02}, in order for bars
to evolve, the halo must be able to participate in the angular momentum
exchange process and to respond to the disc evolution. Thus, in
the simulations the halo must be live, i.e. composed of
particles. Rigid haloes, i.e. haloes which are simply a non-evolving 
potential imposed on the disc particles, artificially hinder or quench
bar formation. }.   
 
AM02 introduced and described two different types of bars, which they
termed MD and MH, respectively. MD-type bars grow in galaxies in which the disc
dominates the dynamics in the inner parts, i.e. are maximum disc
galaxies with haloes with extended cores. On the other hand, MH-type
bars grow in galaxies in which the halo has a small core, i.e. is
centrally concentrated, so that the halo contributes as much as, if not
more than, the disc to the dynamics in the bar region. The 
bars growing in these two types of galaxies have very different
morphology and kinematics (AM02). MH-type bars are thinner and longer
than MD types and their outline is rectangular-like, contrary to MD
bars which have an elliptical-like outline. The amplitude of their $m$
= 4, 6 and even 8 Fourier components of the density reaches a considerable
fraction of the amplitude of their $m$ = 2, contrary to MD-types in which the
$m$ = 6 and 8 are negligible. The density profile along the bar
major axis (face-on view) also differs in the two types 
of bars. In MH-types it has two plateaux, one on each side of the
center, with abrupt drops at the end
of the bar, while in MD-types it drops near-exponentially with
distance from the center (Fig. 5 in AM02). Bars in
MH-type models often have ansae and/or an inner ring, which is
elongated, but not far from circular and has the same major axis as
the bar, as inner rings observed in barred galaxies \citep{buta86}. Their 
side-on 
shape evolves first to boxy and then to a peanut or `X' shape, in contrast
to the MD-types for which the side-on outline stays boxy. The side-on
velocity field of MH-types shows cylindrical rotation, while that of the
MD-types does not. More information on these properties can be found in
AM02. There is a clear connection
between the bar morphology and the amount of angular momentum
emitted by near-resonant material in the bar region and absorbed by
near-resonant material in the halo, in the sense that more angular
momentum has been exchanged in MH-types than in MD-type models
\citep{A05CelMech}. Furthermore, correlations have been found between
the bar strength and the amount of angular momentum absorbed by the
halo \citep{a03}. 

\begin{figure*}
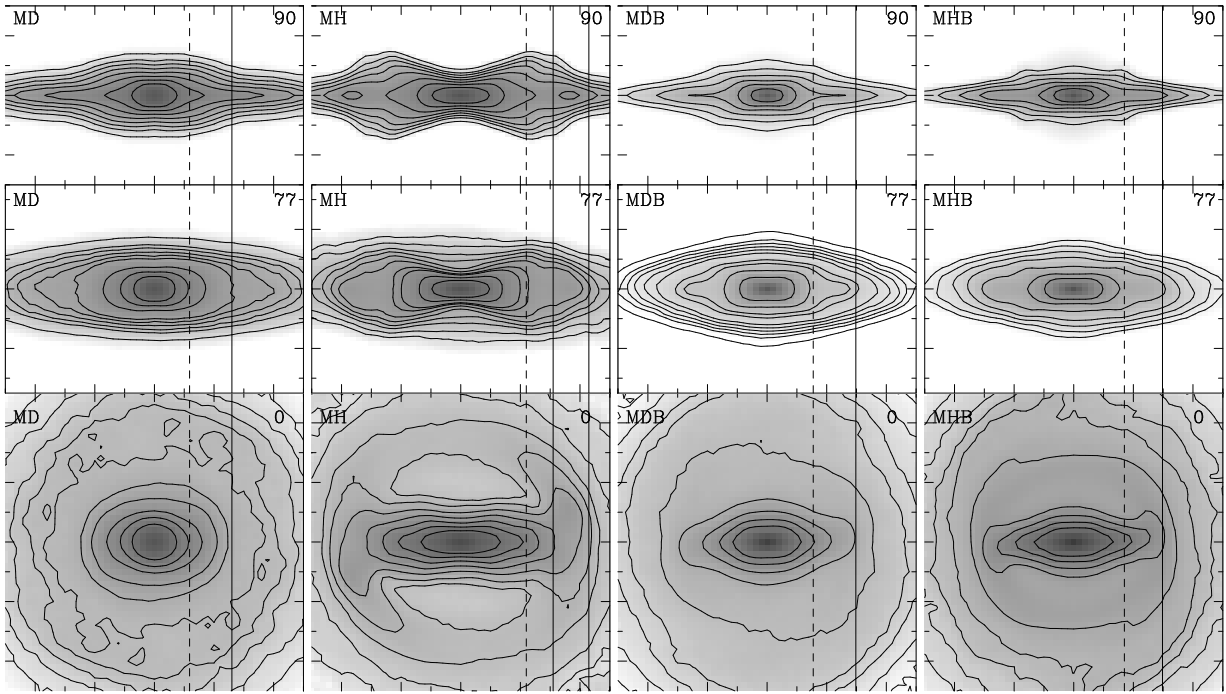

  \setlength{\unitlength}{2cm}
  \includegraphics[scale=0.5,angle=0]{figure1a.ps}
  \includegraphics[scale=0.5,angle=0]{figure1b.ps}
  \includegraphics[scale=0.5,angle=0]{figure1c.ps}
  \includegraphics[scale=0.5,angle=0]{figure1d.ps}
  \caption{Three views of the simulations which will be compared to
  M31. Simulations MD, MH, MDB and MHB are shown in 
  columns 1 to 4, respectively. The upper panels give the edge-on
  views and the 
  lower one the face-on views. The middle ones are views at 
  77$^\circ$, the inclination angle of the M31 disc. The greyscale
  is based on the logarithm of the projected density. The
  isodensities are chosen at levels that show best the shapes
  discussed in section~\ref{subsec:iso}. The solid vertical line(s) in each
  row give(s) a rough
  estimate of the location of the end of the bar, as obtained from the
  face-on view. The dashed lines
  give a rough estimate of the extent of the vertically extended boxy
  feature (i.e. the length of the boxy bulge) as seen from the edge-on
  view. Note in the middle panel that isophotes whose major axis is within 
  the dashed line are boxy-like. On the other hand, isophotes between
  the dashed and the full vertical lines  have a pointed feature,
  elongated roughly in the direction of the bar. This is the signature
  of the vertically thin part of the bar. 
}
  \label{fig:bplength}
\end{figure*}

Since MD and MH-type bars are so distinct, we decided to include one of
each in our fiducial models. Our other two fiducial bars are also
one MH and one MD type, but with a classical bulge component as well. In
the notation of AM02, they are MDB and MHB types. It should be stressed
at this point that, both in observations and in simulations, there is
a continuum of bar types ranging between those described above. Furthermore,
other parameters than the halo mass distribution influence the amount
of angular momentum exchanged within a galaxy and therefore its bar
strength. These include the responsiveness of the halo (in particular its
velocity dispersion), the velocity dispersion in the disc \citep[the Q
  parameter]{ToomreQ} and the existence of a bulge component \citep{a03}. Thus,
it is preferable to consider our four fiducial models as examples of bar
types, rather than as absolute measures of the halo mass
distribution. For simplicity and in order to facilitate
links with previous work, we will refer to our four fiducial models as
MD, MH, MDB and MHB.

The initial conditions of our four models are as described in AM02 and
in \cite{a03}. Thus, the radial density distribution of the disc is
exponential with mass $M_d$ and scalelength $h$ and its vertical
distribution follows a $sech^2$ law with scaleheight
$z_0$. Its radial velocity dispersion is given by the Toomre $Q$
parameter \citep{ToomreQ}. The halo is described by eq. (2.2) of
\cite{hernquist93} and has two characteristic radii, $\gamma$ and
$r_c$, out of which the first one measures the core size and the second
one the outer truncation. The bulge is described by a Hernquist sphere
\citep{hernquist90} 
of mass $M_b$ and scalelength $a$. All the models described here have 
$M_d$ = 1, $h$ = 1, $z_0$ = 0.2, $M_h$ = 5 and $r_c$ = 10. For the
other parameters we have $Q$ = (1.2, 1.2, 1, 1), $\gamma$ = (5, 0.5, 5, 0.5),
$M_b$ = (0, 0, 0.6, 0.4) and $a$ = (-, -, 0.6, 0.4), for models (MD, MH, MDB,
MHB), respectively. Simulations were run using W. Dehnen's treecode
\citep{Dehnen2000:falcON, Dehnen2002a}, as described in AM02 and in
\cite{a03}.  

In these simulation there are about $10^6$ particles in the halo
component and between 200 000 and 400 000 particles in
the visible components (disc and classical bulge). These are
amply sufficient for following the evolution of the run and the formation
of the bar and boxy bulge. For accurate comparisons with  
the observed isophotes, however, one needs isodensities which have less
noise than what is obtained with this number of particles. Rather than
use some smoothing, which could also smooth out real features, we used
the technique described in A05, which allows an $a$ $posteriori$
increase the number of particles. Namely,
we considered ten snapshots, closely spaced in time so that
the bar will not have evolved much in between any two, except of
course for its figure 
rotation. The shapshots were rotated, so that the major axis of the bar
is at the same position angle in all ten cases, and then stacked. 
This increases the number of particles tenfold. We then assumed
two-fold symmetry, i.e. symmetry with rotation by 180$^\circ$ and with
respect to the equatorial plane. This allowed us to avoid asymmetries
inconsistent with M31\footnote{The
  asymmetries in M31 are presumably linked to its two main companions,
  M32 and NGC 205, which are not included in the modeling here.} and
which would, 
therefore, make all comparisons more difficult. Thus, particle numbers
were increased by a factor of 40 at the expense of a considerable
amount of work before each display. This technique allows us to
achieve isodensity contours with very little noise without applying
any convolution which could have smoothed out 
characteristics of the isodensities as well as the noise.    

Face-on and side-on views of the luminous material (i.e. the disc
and, for MHB and MDB, also the classical bulge) can be seen,
respectively, in the lower and the upper panels 
of Fig.~\ref{fig:bplength}, respectively. Comparison of the face-on views
shows that the MD bar is the fattest and shortest of the four and that
its isodensities are elliptical-like. Model MH has the longest and
strongest bar and its isodensities are rectangular-like. For the other
two, the central parts of the isodensities, where the classical bulge
contribution is very strong, are elliptical-like and the parts
further out, where the bar dominates, are rectangular-like. This is
similar to what is observed in many early type strongly barred
galaxies \citep{AMWPPLB}. The MH bar is surrounded by an
inner ring and the MHB bar has clear ansae. 

The side-on views (upper panels) also present many differences. Model MD  
has a boxy-like shape, while MH displays a strong peanut, or `X'-like
feature. It was shown in A05 that in cases with bulges the classical
bulge material `fills' up the central and off the equatorial plane
parts, so that the outline becomes boxy-like. Indeed, models MHB and
MDB have boxy-like shapes.
Structures as those seen in these models are referred to as
boxy/peanut bulges and have been observed in many edge-on galaxies
\citep{ldp00b} and
their link to bars is by now well established (A05 and references
therein). 

Since we can view our simulations both face-on and side-on, it is
possible to obtain estimates of both the bar and of the boxy/peanut bulge
extent. As discussed in AM02, there are several methods for doing this
and their results show often considerable differences. For this reason,
we adopted here very simple eye estimates. We obtained the bar length
from the face-on view by finding by eye the extent of the last
isophote which outlines the bar. The boxy bulge length was measured
as a distance from the center and along the major axis on the side-on
view. Within the boxy bulge extent, the distance  of the isodensities from the
equatorial plane either stays roughly constant (boxy bulges), or
increases with increasing distance from the center (peanut, or `X'
shape bulges). Beyond the boxy bulge extent, the distance  of the
isodensities from the equatorial plane decreases with increasing
distance from the center. Obviously there is some uncertainty in these
determinations. However, if sufficient care is taken, these
measurements can be as reliable as those based on other methods. 
The bar and the boxy/peanut bulge extent are given in
Fig.~\ref{fig:bplength} by the solid and dashed lines, respectively, and
have been extended through all three panels to allow us to compare the
two extents. Our figure shows clearly that the size of this boxy bulge
is shorter than the bar length and that the part of the bar outside
the boxy bulge region is vertically thin. This is in agreement with
previous simulations and with orbital structure results
\citep*{PatsisSkokosAthanassoula2002} and has been discussed
extensively in A05.  

For model MH we plotted two solid lines, i.e. we give two possible
estimates of the bar length. Indeed, if we consider that the bar stops
where the ring starts, then the shortest of the two barlengths is the
appropriate one. Alternatively, if we consider that the bar continues
until the end of the ring, then it is the longest of the two
barlengths that is appropriate. The two alternatives will be discussed
further in Section~\ref{subsec:denprof}. 

The middle row in Fig.~\ref{fig:bplength} gives an intermediate
orientation, with the bar again along the x axis and a galaxy
inclination of 77$^\circ$, i.e. similar to that 
of M31, so as to allow a first comparison. 
In the innermost region the classical bulge dominates and gives
near-elliptical isodensities. The extent and shape of this region
depends on the 
mass and scale-length of the classical bulge compared to the disc
mass and scale-length. Somewhat further out the boxy bulge
dominates. This is a part of the bar, which when seen
face-on is not distinguishable from the bar (see e.g. bottom panels
of Fig.~\ref{fig:bplength}). When the galaxy is seen edge-on and the
bar side-on (i.e. with the line of sight along its minor axis) the
boxy bulge is seen to swell out of the equatorial plane (upper panels
of Fig.~\ref{fig:bplength}). When seen at an angle similar to
that of M31, this gives a more or less (depending on the model) boxy
outline. For some models this is similar to the outline of the
isophotes in the boxy bulge region in M31, as will be seen in
Sect.~\ref{subsec:iso}. The fact that it is the boxy bulge that
dominates the dynamics and sets the isophote characteristics in this
region led us to term this isophotal region as boxy region. 

In regions further out than the boxy region the isophotal shape
varies significantly from one model to another. For model MD the
isophotes show no conspicuous feature and are rather similar to those
in the boxy region. The isophotes in model MH show a very
characteristic pinching toward the center from both sides of the bar
minor axis (i.e. from above and below the galaxy major axis). This is
similar in shape to the pinching that creates the peanut or the `X'
shape in the edge-on views and is in fact
due to it. The part of the isophote that is near the bar major axis shows
two strong protuberances on either side of the centre. The two
remaining models (MHB and MDB) have in this region  isophotes of
similar shape, so they can be described together. Due to the high
inclination and to the fact that the boxy bulge sticks well out of the
equatorial plane, the boxy shape of
the bulge will still be clearly visible near the minor axis. Near the major
axis, however, there will be no contribution from the boxy bulge
(because of its shorter extent) and one will see directly the outer
part of the bar, which is vertically thin, i.e. does not extend much
outside the equatorial plane. This 
contributes an extension, or elongation, of the isodensities towards
the direction of the galaxy major axis and we will refer to the
region where this is seen as the flat bar region. 

\section{comparison with the NIR photometry of M31}
\label{sec:M31}

\subsection{Comparison of observed isophotes with model isodensities}
\label{subsec:iso}

\begin{figure}
  \setlength{\unitlength}{2cm}
  \includegraphics[scale=0.98,angle=0]{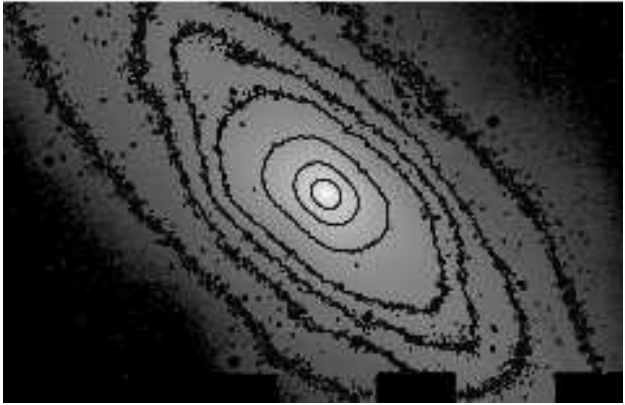}
  \caption{Greyscale representation and isophotes for the $J$ image
    of M31. North is at the top and East to the left.
}
  \label{fig:isophot}
\end{figure}

Fig.~\ref{fig:isophot} displays the observed $J$ image of M31,
presented and analysed in detail in Paper II. The isophotes
show different characteristics in different 
regions of the galaxy. Starting from the center outwards, we first have a
center-most region with elliptical isophotes, presumably due to the
classical bulge. We then encounter a region where the isophotes are
boxy-like (see second and third isophotes from the centre in
Fig.~\ref{fig:isophot}). A similar region was seen in our models in
the previous section and we called it the boxy region. Note, however, that
contrary to the model, these boxy isophotes are skew with respect to 
the galaxy major axis, a feature that we will explain below. Beyond this 
region there is an abrupt change of the isophotal shape (see e.g. the
fifth isophote from the centre in Fig.~\ref{fig:isophot}). Here
two of the sides of the isophotes (the ones near the galaxy minor axis) are
quite straight, similar to the shape of the isophotes in the boxy
region. However, near the major axis, which is at 38$^\circ$, the
isophotes have a clear 
elongation pointing to a direction somewhat, but not much, offset from
the direction of the galaxy major axis. This is similar to what is
seen in two of our fiducial models, namely the two models with a
classical bulge. 
Thus, the shape of the M31 isophotes constitutes a strong argument that
M31 is barred. Let us now examine whether they will allow us to go further
and to set constraints on the bar parameters.

\begin{figure*}
  \setlength{\unitlength}{2cm}
  \includegraphics[scale=0.97,angle=0]{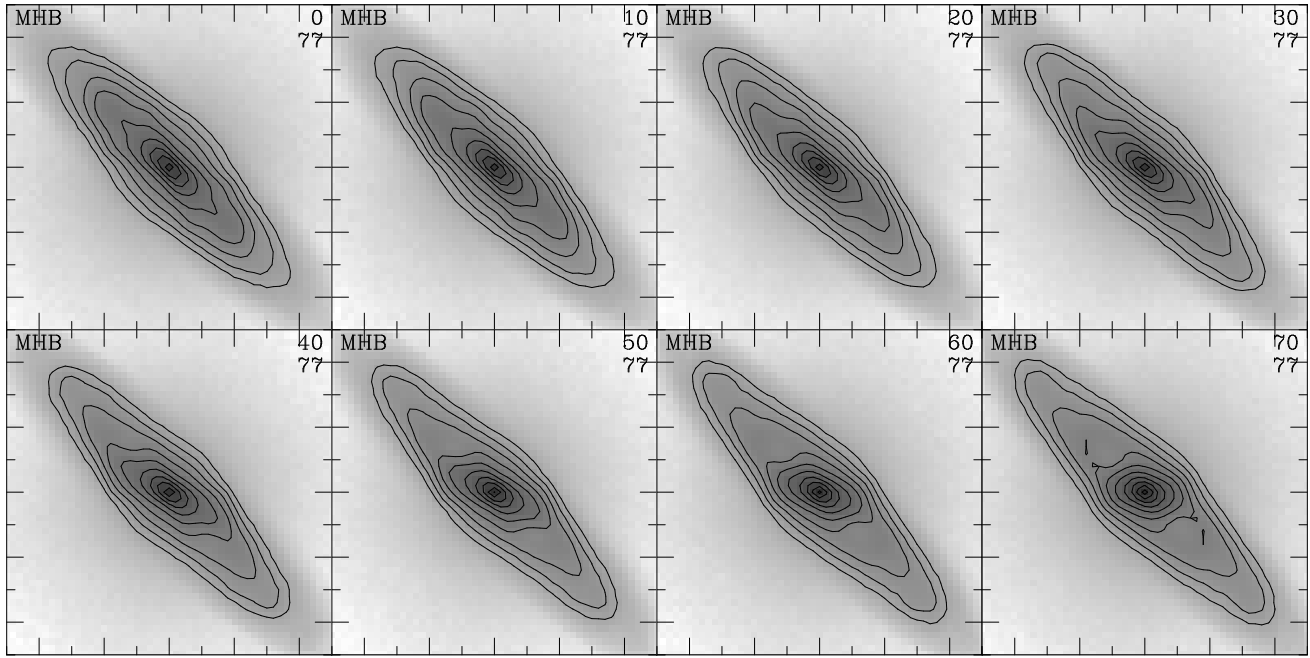}
  \caption{Isodensities for model MHB and various angles between
    the bar and the galaxy major axes. From left to right and top to
    bottom this angle varies from 0 to 70$^\circ$ in steps of 
    10$^\circ$. The inclination angle is in all cases 77$^\circ$.
}
  \label{fig:PA}
\end{figure*}

\begin{figure*}
  \setlength{\unitlength}{2cm}
  \includegraphics[scale=0.98,angle=0]{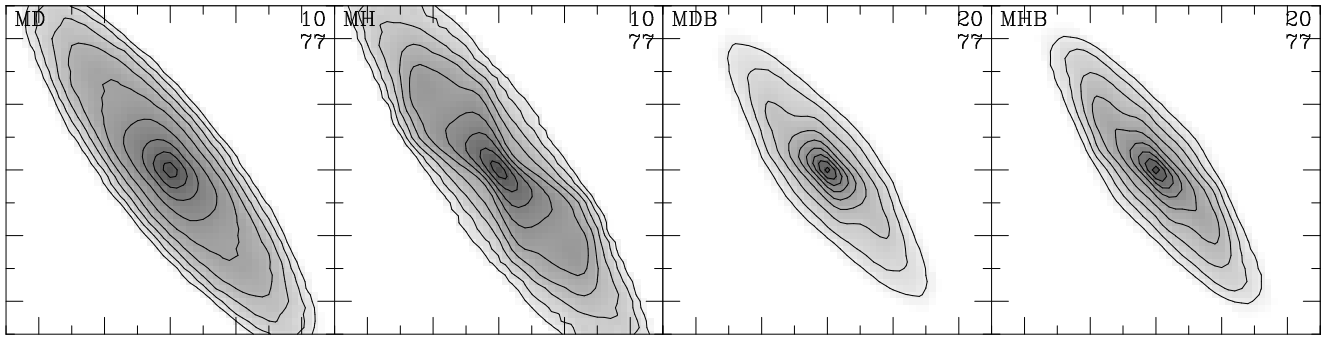}
  \caption{The four fiducial models viewed in a M31-like
  orientation, i.e. an inclination of 77$^\circ$ and a position angle
  of 38$^\circ$. Measured on the plane of the
  galaxy, the angle between the 
  galaxy major axis and the bar major is 10$^\circ$ for the two models in
  the left panels and 20$^\circ$ for the two models in the right
  panels. The isodensities are chosen at levels showing best the
  features discussed in Section~\ref{subsec:iso}. 
}
  \label{fig:comparison}
\end{figure*}

An important feature to note here is that the isophotes of M31 do not
show any pinching towards the center and this allows us to eliminate
very strong bars, such as those in model MH, as likely candidates for
reproducing the isophotal 
structure of M31. Indeed, the horizontal and vertical
structures of a bar are strongly coupled. As shown by the simulations
of A05 (see also AM02), the strongest bars have clear peanut (or
even `X') shapes when viewed edge-on and side-on, while the less
strong bars have  
a boxy side-on shape. This is in good agreement with the observational
results of \cite{ldp00b}, who showed that
stronger bars have more prominent peanut shapes. Peanuts, however,
and even more so `X' shapes, have a narrowing of the
isophotes/isodensities towards the center if there is no bulge
component and this is preserved also
for viewing angles not far from edge-on. Thus, strong bars
viewed at such angles would always show this characteristic squeezing
or pinching of the isophotes in the central parts and the fact that this is
absent in M31 argues that its bar can not be very strong. 
The isodensities of the MD model also do not reproduce well the
observed isophotes, since, in the flat bar region, they miss the
characteristic elongation in a direction near to the galaxy major
axis. Thus, MD-type bars are also inconsistent with the M31
bar. In fact, only the two models with a classical bulge, MHB and MDB,
give a good representation of the isophotes in all the relevant
region. Even those, however, do not reproduce the slight
skewness of the isophotes (i.e. their asymmetry with respect to the
galaxy major axis) in the orientation shown in the middle panels of
Fig.~\ref{fig:bplength}. This, nevertheless, can be achieved if we
consider a different viewing angle.
 
Three angles are necessary in order to determine the orientation of
the M31 disc and bar, namely the inclination, the angle between the bar major
axis and the galaxy major axis and the position angle of the
galaxy. This third angle does not contain any physical information,
and will not affect the comparison of isodensity and isophotal
shapes. The constraints brought by our simulations to the inclination angle
are not tight. For one thing, there is no way of distinguishing
between front and back, since $N$-body models do not contain
dust. Moreover, the axial ratio of the inner isophotes is a function of
three things : the inclination angle of the galaxy, the bar axial
ratio in the galaxy equatorial plane (since a fatter 
bar will need to be viewed nearer to edge-on than a thinner one to give
the same projected axial ratio) and the vertical extent of the bar 
material. Thus, the observed axial ratio in the inner regions can not
determine uniquely the inclination angle. 

In Fig.~\ref{fig:PA} we vary, for model MHB,  
the angle between the bar and the galaxy major axis, to check its
effect on the isodensity shapes. When this angle
is zero, i.e. when the bar lies on the galaxy major axis, all
isodensities are symmetric with respect to this common axis, as
expected. This symmetry is broken as soon as this angle becomes
nonzero. Then the boxy isophotes in the inner parts become skew, while
the elongation of the isodensities somewhat further out is offset from
the galaxy major axis. Both these features are seen in the
corresponding isophotes of M31, thus arguing that the bar in this
galaxy is at an
angle with the line of nodes. The direction of the skewness of the boxy
isophotes and the direction at which the tips of the isophotes in the
flat bar region
are pointing gives an indication about the position angle of the
bar. Both directions are pointing further to the SE of M31's northern semi-major
axis and this tells us that the position angle of the bar major axis
is larger than the position angle of the galaxy. The amount of
skewness and the corresponding isophotal shapes give us some constraints
on the difference of these angles, as can be seen in
Fig.~\ref{fig:PA}. The best fits are for 20$^\circ$ and 30$^\circ$
(measured on the plane of the galaxy).
Indeed, for 0$^\circ$ and 10$^\circ$ the asymmetries and skewness are  
insufficient, while angles of 40$^\circ$ and larger 
again do not give a good representation. It should be noted, however,
that the exact values of this range may depend on the model. 

Fig.~\ref{fig:comparison} uses appropriate orientations for all
four models, to allow a full comparison with the M31 isophotes. It
shows that, even with an optimum angle between the bar and the galaxy
major axis, models MD and MH do not fare well. On the other hand,
models MDB and MHB reproduce well the characteristic of the M31
isophotes, including the observed skewness. Thus, comparing isophotal and
isodensity shapes has given us precious information on the existence
and the properties of the M31 bar. 
  
\subsection{Radial luminosity and density profiles}
\label{subsec:denprof}

In this section we compare radial luminosity profiles made along cuts on
M31 as projected on the sky and radial density profiles obtained from
the models in a similar way.

\begin{figure}
  \setlength{\unitlength}{2cm}
  \includegraphics[scale=0.4,angle=0.]{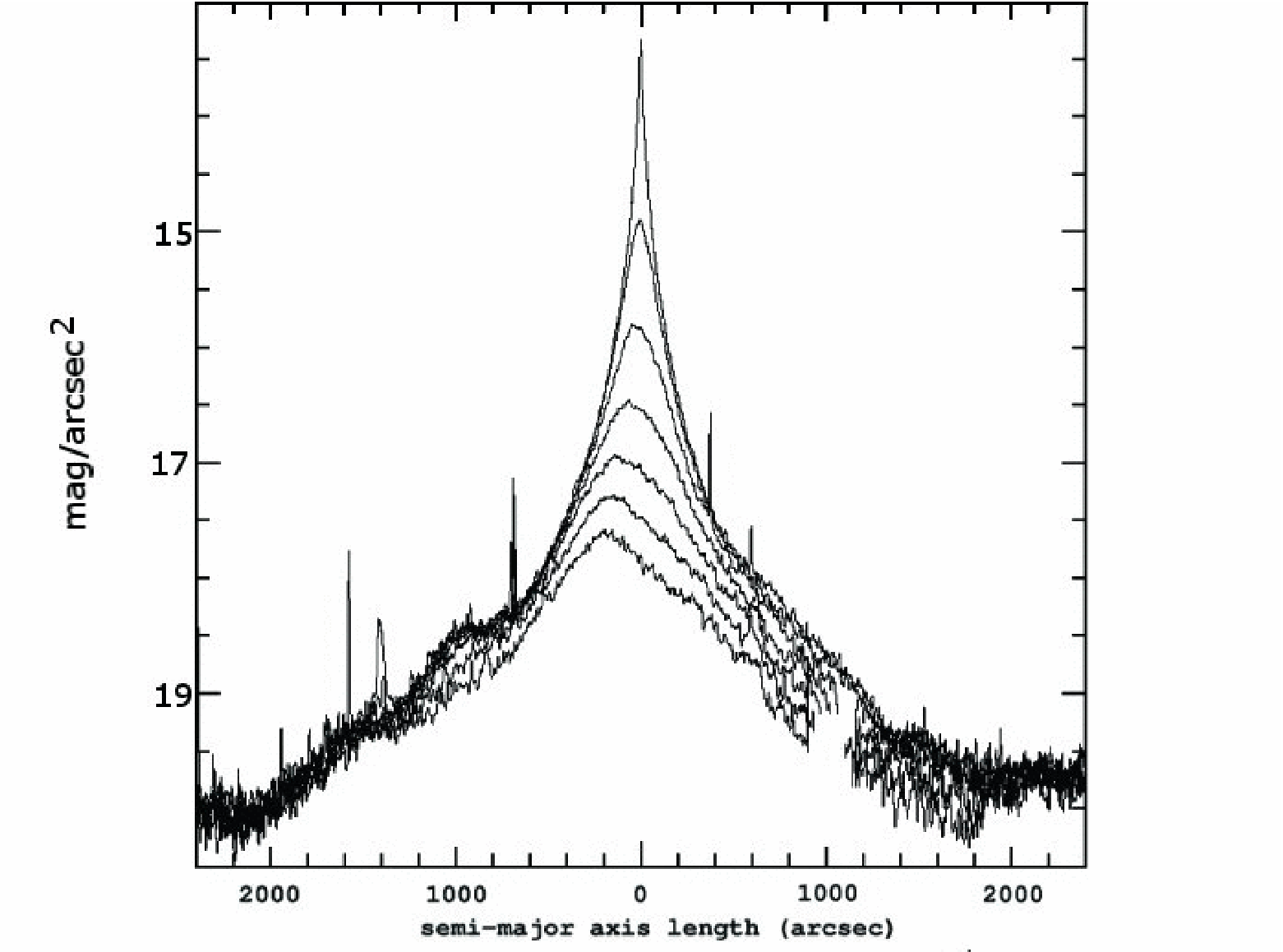}
  \caption{Luminosity of M31 along cuts parallel to the galaxy major axis and
    offset from it towards the SE by $0''$, $50''$, $100''$, $150''$, $200''$,
    $250''$ and $300''$. The distance along the cut is measured on the
    abscissa in arc seconds. The ordinate is labeled in $mag/arcsec^2$,
    to within an arbitrary constant. Note the characteristic asymmetry
    of the luminosity profile with respect to the center. 
}
  \label{fig:M31cutsS}
\end{figure}

\begin{figure}
  \setlength{\unitlength}{2cm}
  \includegraphics[scale=0.4,angle=0.]{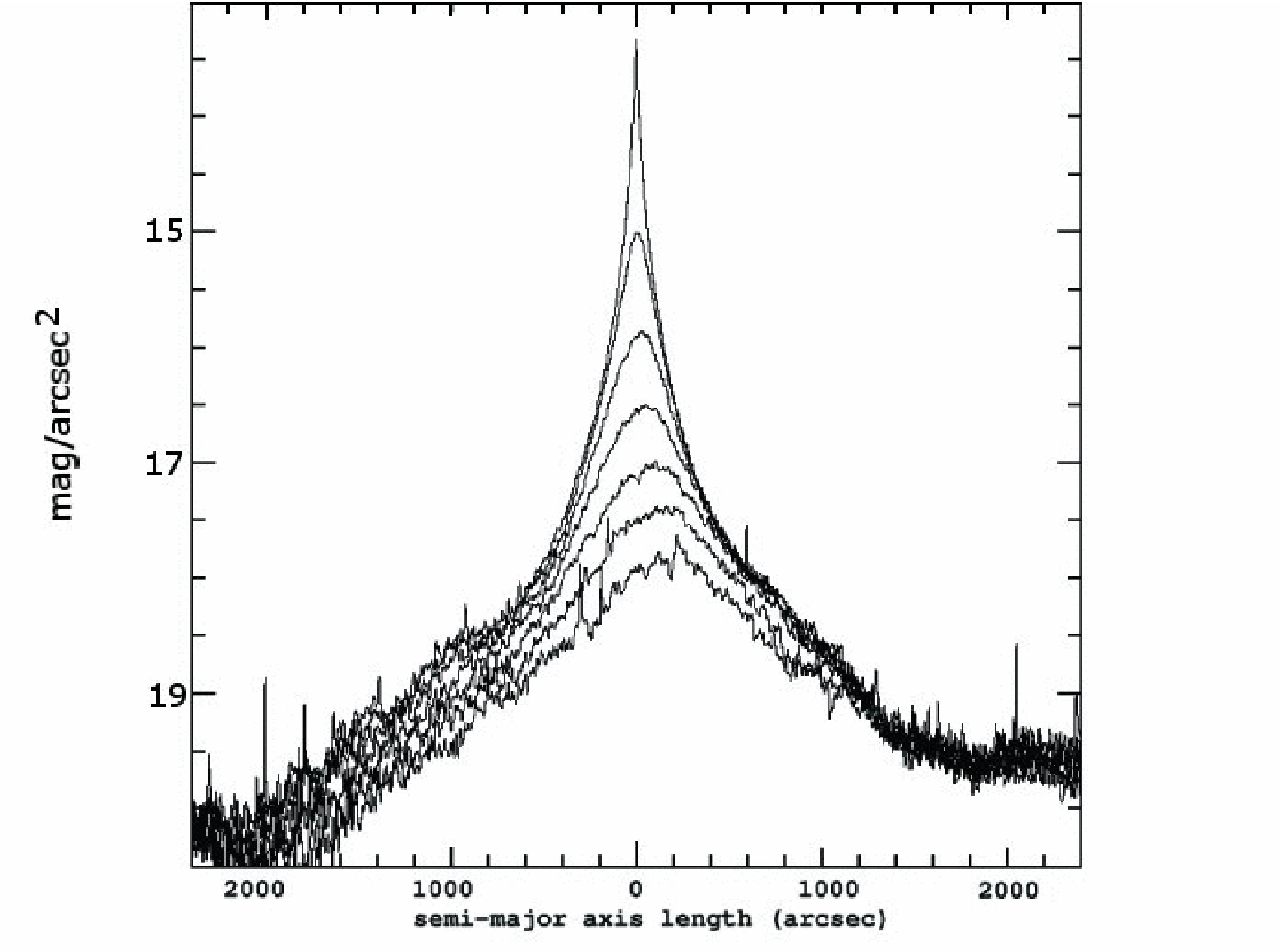}
  \caption{As in Fig.~\ref{fig:M31cutsS}, but for cuts offset towards
    the NW.  
}
  \label{fig:M31cutsN}
\end{figure}

Using the $J$ image of Paper II, we obtained the radial luminosity
profile on a cut along the galaxy major axis, as well as on cuts
parallel to it but offset by multiples of $50''$ either to the SE or
to the NW. The results are given in
Figs.~\ref{fig:M31cutsS} and \ref{fig:M31cutsN}, for the SE and the
NW, respectively. In all profiles we note interesting asymmetries
between the two sides. First, the position of the maximum drifts with
respect to the center with increasing offset. For offsets to the NW,
the maximum drifts towards the SW, while for offsets to the SE it
drifts towards the NE. Note also that on the SE cuts there is a hump
on the profiles to the NE of the center. In fact, one can see there a
plateau ending at about $1000''$ from the nucleus, followed by a
relatively steep drop starting at $1000''$ and ending roughly at
$1300''$. The intriguing thing here is that this plateau occurs only
on the one side of the centre. Such an 
asymmetry is also seen on the cuts parallel to the galaxy major axis
and offset to the NW of the galaxy (Fig. \ref{fig:M31cutsS}). In this
case, however, the bump is on the SW side of the nucleus and is less
prominent. 

\begin{figure*}
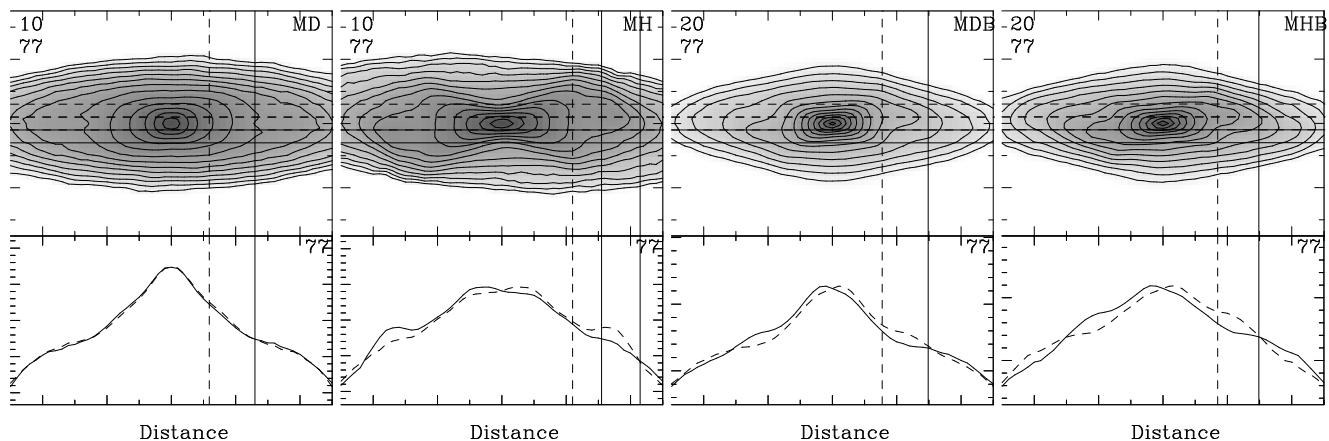

  \setlength{\unitlength}{2cm}
  \includegraphics[scale=0.54,angle=0]{figure7a.ps}
  \includegraphics[scale=0.54,angle=0]{figure7b.ps}
  \includegraphics[scale=0.54,angle=0]{figure7c.ps}
  \includegraphics[scale=0.54,angle=0]{figure7d.ps}
  \caption{The upper panels show the luminous material of our four
    fiducial simulations, viewed at an inclination angle of
    77$^\circ$. For the two models on the left, the angle between the
    bar major axis and the galaxy major axis, measured on the plane of
    the galaxy, is 10$^\circ$. For the two models on the right it is
    20$^\circ$. The
    horizontal stripes delineate the regions from which we calculate the
    density. The lower panels show the logarithm of the density as a
    function of 
    position along these stripes. The solid line corresponds to the
    stripe outlined with the solid horizontal lines in the upper
    panels, and the dashed 
    one to the stripe outlined with the dashed horizontal lines. The
    vertical lines give the lengths of the bar and of the boxy bulge,
    as found from Fig.~\ref{fig:bplength} if no projection is taken
    into account. 
}
  \label{fig:modelcut}
\end{figure*}

The existence of the humps and the asymmetry of their location on the
cuts provide useful clues about the existence and the properties of
the bar in M31. Indeed, the observations of \cite{Elm**2}
show such humps in similar cuts in strongly barred early type
barred galaxies. In that paper, the cuts were made along the bar major
axis and show 
that the humps are symmetrically located with respect to the
nucleus. Such humps are also seen for model MH in AM02 (see their
Fig. 5) and they are also symmetrically located on either side of the
center. On the other hand, no such humps are seen in most late type
barred galaxies \citep{Elm**2}, where the decrease of
the density is exponential-like. A similar decrease is found for model
MD in AM02 (their Fig. 5). This could be a clue that the M31 bar is
more likely to be MH-type than MD-type.

Let us now follow the clues given by the asymmetric location of the
bumps. Humps alone are not unambiguous evidence of the existence of a bar,
since they can also be made by a lens-like
structure with abrupt edges, even if the lens is circular or
near-circular. Such a structure, however, could not be responsible for
the M31 humps, since any axisymmetric, or near-axisymmetric, structure
would place the humps symmetrically with respect to the nucleus. Thus, the
existence and the asymmetry of the humps, when taken together, is an
argument for the existence of a bar in
M31. If the major axis of this bar coincided with the galaxy major
axis, the humps would again be located symmetrically with respect to
the nucleus. Thus, the asymmetry of the bump locations argues
that they are due to a strongly non-axisymmetric feature, i.e. a bar,
whose major axis is at an angle with respect to the galaxy major
axis. This result is in good agreement with was found in the last
section by comparing isophotes to isodensities. 

The location of the bumps can also give indications on the angle
between the galaxy and the bar major axes, as well as on the length of the
bar. We show this for our four fiducial models in
Fig.~\ref{fig:modelcut}. For models MH and MD we rotated the disc so
that the left (eastern for comparison with observations) part of bar
major axis 
is at an angle of 10$^\circ$ to the left (eastern) part of the galaxy
major axis and then projected it to an inclination angle of 77$^\circ$. We
did the same thing with models MDB and MHB, except that now we chose
the angle between the bar and the galaxy major axis to be 20$^\circ$.
We used 10$^\circ$ and 20$^\circ$, respectively, because isophote to
isodensity comparisons (Sect.~\ref{subsec:iso}) show that these angles
best reproduce the M31 observations. Other values of these angles
will be discussed later in this section. 
The results are given in the upper panels of  
Fig.~\ref{fig:modelcut}. We then make cuts parallel to the galaxy 
major axis with offsets above and below. The density along these cuts
can be seen in the lower panels of Fig.~\ref{fig:modelcut}. 
We find that the cuts for model MD are symmetric with respect to the
center, the maxima in the 
two stripes coincide and there are no humps on either of the
cuts. This argues strongly that an MD type model is a very unlikely
candidate for the M31 bar. The cuts for model MH show an asymmetry,
the maxima in the two stripes are offset and there are humps
asymmetrically located, as in M31. The amplitude of these humps,
however, is too large and thus this model is not a good
candidate for M31. On the other hand, models MHB and MDB fare much
better. Their profiles are asymmetric and in the same way as
M31. In particular, the position of the maximum shifts from the center
in the same direction as for M31, i.e. towards the left (eastern) part
for cuts to the South and towards the right (western) part for cuts to
the North. Furthermore, we note the existence of a single hump on each
profile, 
located in the direction towards which the maximum of the profile
is shifted, again in agreement with what is observed in M31. 
This argues that, as in the MHB and MDB models in Fig.~\ref{fig:modelcut},
the eastern semi-major axis of the bar in M31 is to the South of the
eastern galaxy semi-major axis. If it were to the North, then the
asymmetries would be reversed. Thus, the sense of the asymmetries
on the radial luminosity profiles give us constraints on the way the
bar is oriented with respect to the galaxy major axis.

Such cuts give us information not only on the sign of the angle, but also
constraints as to its value. We repeated the above analysis considering
different angles between the bar and the galaxy major axis in steps of
10$^\circ$ (results
not shown here). For model MD we find that no angle gives a reasonable
fit. For model MH, the best fit is for an angle of 10$^\circ$, but, as
can be seen in Fig.~\ref{fig:modelcut}, even that is not good. For
model MDB we find that the best fit comes for an angle of 20$^\circ$,
but 30$^\circ$ is still good. Finally, for model MHB both 10$^\circ$
and 20$^\circ$ are good. Thus, we can conclude that the most probable
angle is about 20$^\circ$. This is in good
agreement with the range of values we found in the previous
section. Adopting the M31 disc position angle 
of 38$^\circ$, this means that the position angle of the M31 bar is of the
order of 45$^\circ$. 

Further information can be obtained on the length of the bar. For this
we plotted in Fig.~\ref{fig:modelcut} vertical lines at the radii at which we
estimated the length of the bar and of the peanut component from the
face-on and end-on views, respectively
(Sect.~\ref{sec:simul}). Note 
that the bar length is beyond the end of the flat part of the hump,
nearer to the end of the abrupt drop after the plateaux. Using this,
it is possible to make a rough estimate of the length of the bar in M31. 
As seen in Fig.~\ref{fig:M31cutsS}, the flat
part of the bump there extends till $1000''$ from the center, while the
abrupt drop after the flat part extends from $1000''$ to $1300''$.
Thus, we can estimate the length of the bar in M31 to be of the order
of $1300''$. 
Of course this estimate has a considerable uncertainty, since we can
not determine the end of the steep drop in the profile precisely either in the
model, or in the data. There is a further uncertainty from the fact
that the bar semi-major axis is at an angle to the galaxy semi-major
axis, which we have only roughly constrained. Changes of the order of
2\%, 5\% or 14\%  are to be taken into account if this angle is
10$^\circ$, 20$^\circ$ or 30$^\circ$, respectively. 
 
For model MH we plotted two vertical lines for the bar length,
corresponding to the two estimates of the bar length given in 
Fig.~\ref{fig:bplength}. The shorter of the two is obtained if we
consider that the bar ends where the ring starts and the longer one if
we assume that the bar continues within the ring reaching its outer
end. From Fig.~\ref{fig:modelcut} it is clear that the end of the
sharp drop after the end of the plateau coincides with the second
definition, giving the longer barlength. This, however, does not mean
that it is this definition that is correct, since clearly the
ring participates in the formation of the hump.

\section{Kinematics}
\label{sec:kinematics}

The kinematics of M31 have been well observed, both in the HI and the
H$\alpha$. 
Our $N$-body models, however, do not contain
gas, so it is not possible to make any direct comparisons to the
corresponding observations. Nevertheless, the observed gaseous
kinematics give strong arguments for the existence of a
bar in M31. For this reason, we will discuss here first the
relevant observational results and then present simulated PVDs of the
gas component in a strongly barred galaxy model from \cite{A92a}. 
Although this is an idealised model (with a Ferrers \citeyear{ferrers}
  ellipsoidal bar)
and not a simulation result as the four fiducial models we discuss
here, it will, nevertheless, 
allow us to discuss some generic features of PVDs in barred galaxies.

\begin{figure}
  \setlength{\unitlength}{2cm}
  \includegraphics[scale=0.33,angle=-90.]{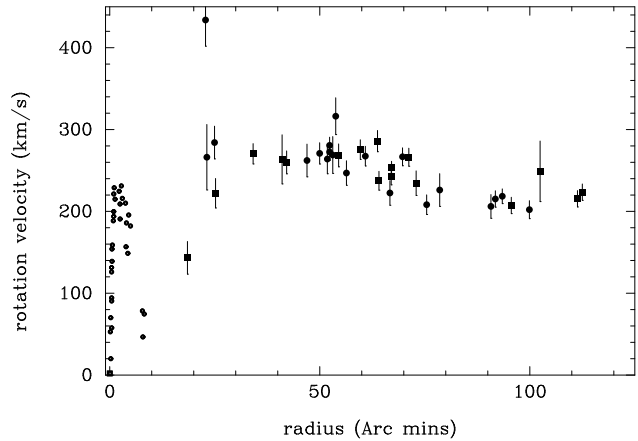}
  \caption{Rotational velocities of M31 from Rubin \& Ford (1970).
    Note the deep minimum around $10'$.
}
  \label{fig:M31RF}
\end{figure}

The first clear kinematical clue that M31 is barred was contained in
the data of \cite{RF} but, at the time, was not interpreted as such. The
velocity measurements from that paper, read off their Fig. 9, are
reproduced in Fig.~\ref{fig:M31RF}. They show a very deep minimum near
$10'$, which is not observed normally in spiral galaxy rotation 
curves. This minimum, described as `the annoying deep minimum' by
\cite{Rubin}, provoked discussion, often going as far as to 
cast doubts as to the quality of the data. 

\begin{figure}
  \setlength{\unitlength}{2cm}
  \includegraphics[scale=0.5,angle=0.]{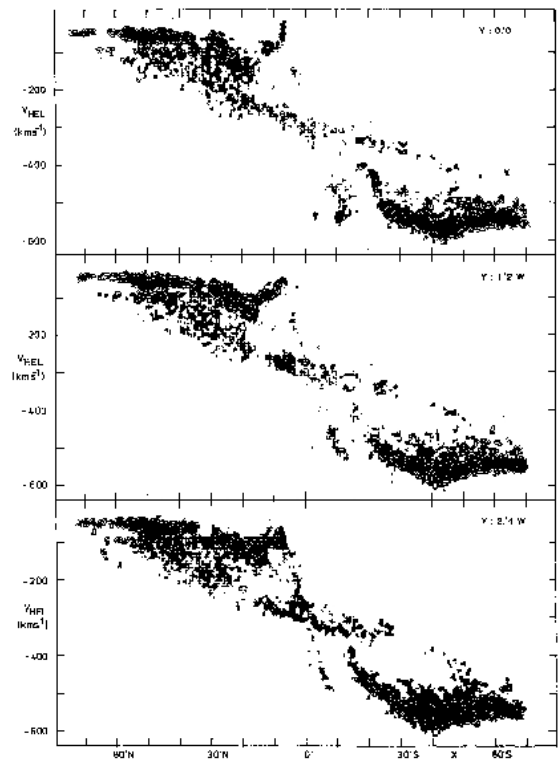}
  \caption{Observed PVDs from the HI in M31 from cuts parallel to and
    near the galaxy major axis. Reproduced from Brinks \& Shane (1984). 
}
  \label{fig:M31BS}
\end{figure}

Brinks \& Shane (1984) observed M31 in HI and gave a number of PVDs
on cuts parallel, or perpendicular to the major axis. The cuts
parallel to and near the major axis
(some of which are reproduced in Fig.~\ref{fig:M31BS}) show very
interesting structure. In particular, they show two branches,
separated by a low density region. In order to explain and reproduce
these features, \cite{BrBu}
proposed a model of a rotating disc which flares and warps in its
outer regions. According to their model, the inner branch would be due
to gas near the galaxy major axis, which would imply that the rotation
curve for that galaxy would be very sharply rising. It would then be
the outermost gas which, projected onto the line of sight, is
responsible for the lower branch (the one where the
velocities rise slower with distance from the center). Their model
reproduces a number of the main 
characteristics of the PVD diagrams (namely the existence of two
branches through the center), but can not be qualified as a good fit to the
data. 

\cite{Rubin} superposed the \cite{RF} velocity measurements
on the \citeauthor{BrSh} PVDs going through the center of the galaxy and
found good agreement. She  
thus established that there was nothing wrong with the \citeauthor{RF}
data and stressed that the low velocities should not be identified  as
rotational velocities. She also proposed that the
complex velocities could be due to extra-planar gas in the innermost
parts, without, however, showing that any such model could indeed
reproduce the PVD characteristics. Thus, both \citeauthor{BrBu} and
\citeauthor{Rubin} relied on 
extra-planar gas in the central and/or in the outer parts of the
galaxy, but did not invoke the possibility of a bar. It was only a few
years later that it became clear that such structures arise naturally
if the galaxy is barred.

\begin{figure}
  \setlength{\unitlength}{2cm}
  \includegraphics[scale=0.4,angle=-90.0]{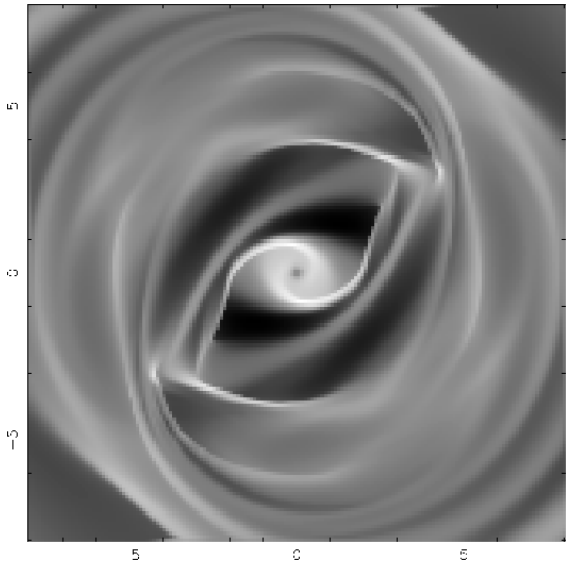}
  \caption{Response of the gas to a bar. The major axis of the bar is
    at 45$^\circ$ to the horizontal and is rotating clockwise. It has
    a semi-major axis length 
    of 5 and a semi-minor axis of 2 length units. Lighter shades
    denote higher density regions.  
}
  \label{fig:gasresp}
\end{figure}

\begin{figure*}
  \setlength{\unitlength}{2cm}
  \includegraphics[scale=0.6,angle=-90.0]{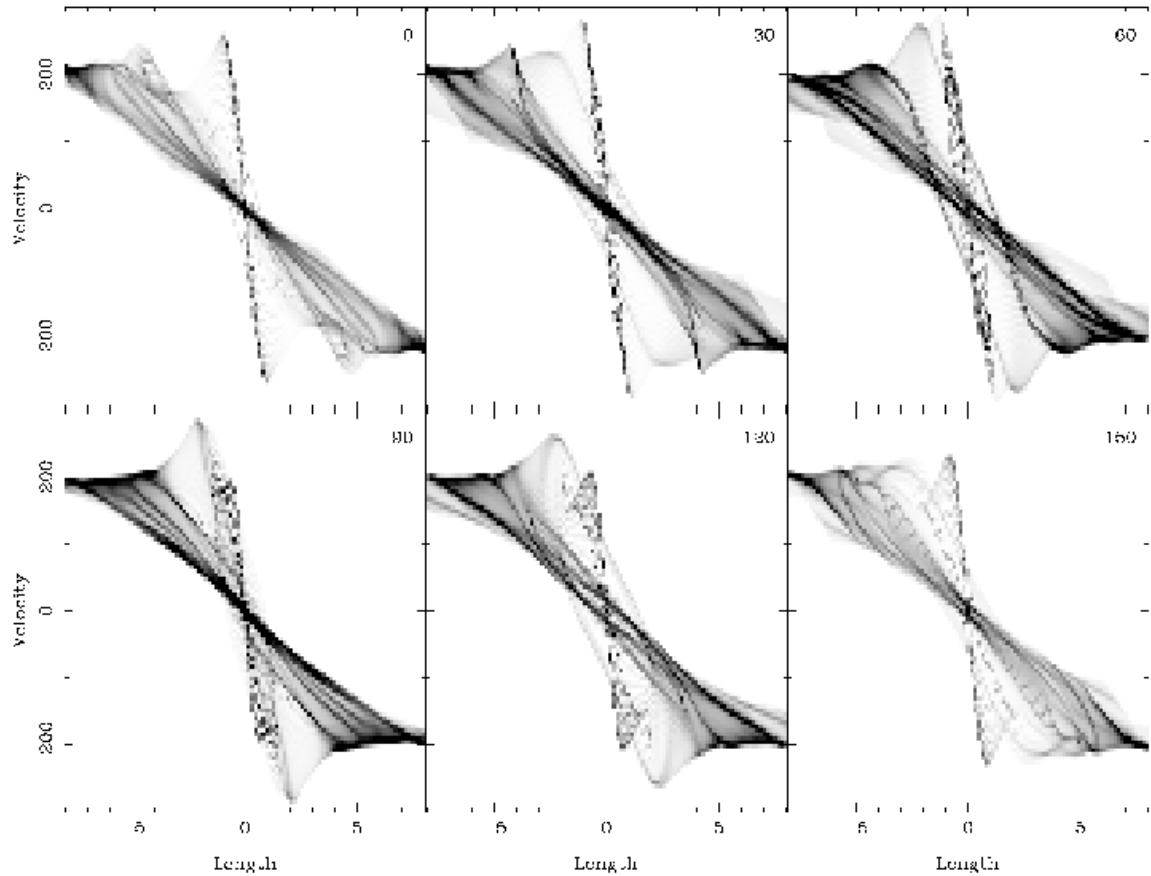}
  \caption{Model PVDs for the gaseous component in the strongly barred
    galaxy used in Fig.~\ref{fig:gasresp}. Results are presented for
    six values of the angle between 
    the galaxy and the bar major axis (given in the upper right
    corner of each panel). 
}
  \label{fig:PVD1}
\end{figure*}

Structure such as seen in the above discussed PVDs has been witnessed
in the observed emission line PVDs of edge-on galaxies with boxy or
peanut bulges and was readily attributed to the bar \citep*{km95, 
mk99, bf99}. Indeed,
gas responding to a strong bar potential does not fill the bar region
uniformly, but, on the contrary, forms strong concentrations and
relatively empty regions \citep*{A92b}. This is illustrated in
Fig.~\ref{fig:gasresp}, which shows the gas response to a strong bar
model (model 047 of
\cite*{A92a}). We see that most of the bar region is relatively empty
of gas except for the region surrounding the center and two narrow stripes
along the leading edges of the bar. These are in fact the loci of
shocks in the gas flow \citep*{A92b}. Secondary concentrations can be
seen surrounding the bar region and at the ends of the bar. As shown
by \cite{ab99}, it is the 
existence of these high density and low density regions, together
with the corresponding velocities, that creates
the characteristic structures of the PVDs in barred
galaxies. To show this, \cite{ab99} obtained model PVDs by projecting
edge-on some of the two-dimensional hydrodynamic simulations of
\cite*{A92b}. We
repeat this for the model in Fig.~\ref{fig:gasresp} and display the
result in Fig.~\ref{fig:PVD1}.  One can immediately see the two
branches that stand out in the M31 PVDs, 
separated by a low density region. The inner branch 
is due to the gas near the center, where the flow lines are elongated
roughly along the bar minor axis. The outer and lower of the two
branches is due to material in the 
outer regions. The two branches are separated by low density regions on the
PVDs, because the two regions are separated by low density regions in
the face-on view of the bar. 

Further comparison between Figs.~\ref{fig:M31BS} and \ref{fig:PVD1}
allows us to check whether the values for the bar position angle and
length, as obtained from the photometry, are consistent with the
observed PVDs. In the previous sections we found the best fit
between models and observations when the angle of the bar major axis is
around 20$^\circ$ from that of the galaxy major axis. This is near the
viewing angle of the 
upper middle panel in Fig.~\ref{fig:PVD1}. We note that in this
case the contribution from the central part, i.e. the inner branch
of the PVD, is a nearly straight segment passing through the center of
the PVD, in agreement with the observations. Also, this segment rises to
velocities higher than those of the outer parts, again in agreement
with the observations\footnote{As explained in \cite{ab99}, this is
  due to the fact that in the central regions the gas flow lines follow
  roughly the orientation of the x$_2$ family. The largest velocities
  occur of course near pericenter and would, in this region, be
  observed when the bar is viewed near side-on, as is the case.}. Note
that this latter check excludes a large value of 
the angle between the bar and galaxy major axes, as e.g. in the lower
middle panel of Fig.~\ref{fig:PVD1}. A further consistency check can be made
with the bar length. From Fig.~\ref{fig:PVD1} it can be seen that the
maximum extent of the low density region between the two branches of
the PVD is somewhat smaller than the length of the bar semi-major
axis. Applying this to Fig.~\ref{fig:M31BS} we see that the barlength
of $22'$, which we estimated in Sect.~\ref{subsec:denprof} based on the
photometry, agrees with the observed PVDs.

Unfortunately, the analysis of the kinematics can not be pushed any
further than the discussion of generic bar features for many
reasons. The bar model used in this section is an idealised bar model
and not one of the four fiducial models discussed in the previous
sections. Moreover, the PVDs of Fig.~\ref{fig:PVD1} 
were obtained by integrating along the line of sight a perfectly
edge-on razor thin gas layer, whereas the inclination angle of M31 is
77$^{\circ}$, its gas layer is not razor-thin and this layer could also flare
and warp in the outermost parts. Although this is far outside of the
established bar 
region, it could still influence the PVDs if gas was folded onto the
line of sight. This would give further branches on the PVDs. 
Indeed, the M31
PVDs \citep[see Fig. 8 of ][]{BrSh} show suggestions of many
more features which, due to the lack of gas in extended areas of M31,
can not be followed clearly. For example, in many of the
cuts one has the distinct impression that there are {\it three}
branches. To model all these features fully 
is an interesting, but daunting task, further complicated by the
lack of gas in some crucial regions, and well beyond the scope of this
paper.   

\begin{figure}
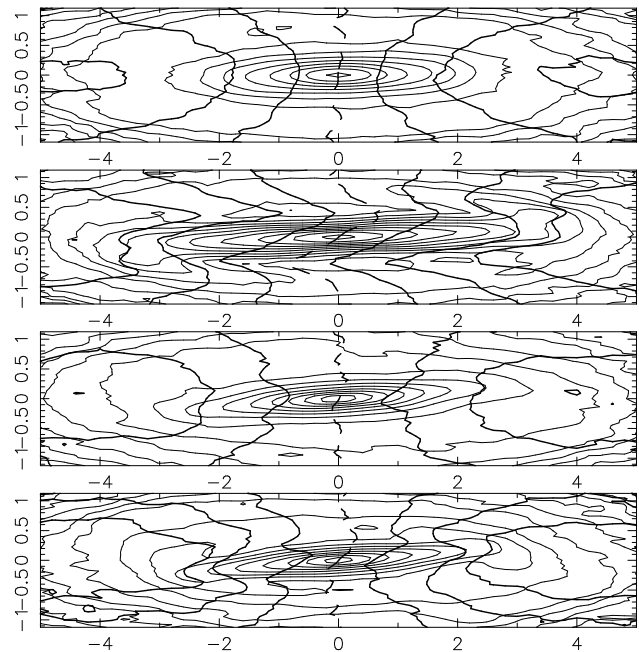

  \setlength{\unitlength}{2cm}
  \includegraphics[scale=0.5,angle=0.0]{figure12a.ps}
  \includegraphics[scale=0.5,angle=0.0]{figure12b.ps}
  \includegraphics[scale=0.5,angle=0.0]{figure12c.ps}
  \includegraphics[scale=0.5,angle=0.0]{figure12d.ps}
  \caption{Stellar velocity field of the four models discussed in this paper,
    namely, from top to bottom, models MD, MH, MDB and MHB. The
    isovelocities are given by thick lines and the kinematic major
    axis with a dashed line. We also overlay the isodensities with
    thin contours. 
}
  \label{fig:stel-vel-field}
\end{figure}

The stellar velocity field of our four models is given in
Fig.~\ref{fig:stel-vel-field}. Unfortunately, no corresponding
observational data is available, so these figures can only be
considered as predictions. For the viewing angles we adopted the
values found to be optimum in the previous sections, i.e. an angle
between the bar and galaxy major axes of 10$^\circ$ for models MD and
MH and to 20$^\circ$ for models MDB and MHB. The inclination angle was
taken to be 77$^\circ$ in all cases. When calculating these velocity fields 
we included both the disc and, whenever relevant, the classical bulge
material. This makes the effect of the 
bar less strong, but will allow comparisons with future observations. 
 
The velocity field of model MD shows no clear-cut perturbations which
could serve as clues 
to the existence of the bar. On the other hand, the isovelocities of
model MH have sharp deviations due to the bar, which give them the
form of a letter `Z', as seen in many barred galaxies (e.g. \cite{PRFT}). Such
deviations, but less clear cut, are seen also for models MDB and
MHB. A telltale sign for the existence of a bar or of an oval is the
fact that the kinematical and photometric minor axes do not align 
\citep{bosmat,bosmap}. This would have been easily spotted for models
MH, MDB and MHB, as well as in a very careful analysis of the velocity
field of model MD.

\section{Discussion}
\label{sec:discuss}

\subsection{Comparison with previous work}
\label{subsec:comparison}

Looking at the isophotes of M31 \citep[][etc]{lindblad, deVauc}
one immediately sees that the part considered as the bar in
previous work \citep{lindblad, Stark, StarkBin}
is in fact the boxy/peanut bulge and in the last two papers was often
termed as such. Thus, the `bar' they propose should
be shorter than ours, and also its angle with the northern 
semi-major axis of M31 should be larger. Indeed, this is the case.
In our model the bar semi-major axis length is about $22'$, the
angle between the galaxy and the bar semi-major axes is 
about 20$^\circ$ and the axial ratio in the equatorial plane larger
than 2. In the previous models the `bar' semi-major axis
length is about $15'-16'$, the angle between the major axes of the
galaxy and the `bar' is about 70$^\circ$ and the axial ratio in the
equatorial plane smaller than 2. These latter numbers are in
agreement with those of the
boxy bulge of our model. A more precise comparison is not possible
since the 2D shape of our boxy/peanut bulge is rectangular-like, while
that assumed by \cite{lindblad}, \cite{Stark} and \cite{StarkBin}
is an ellipsoid.
 
\subsection{Spiral Structure}
\label{subsec:spiral}

The most prominent feature in M31 is a ring around $50'$. More precisely,
this is a pseudo-ring, since it has an opening in the SW and its
distance from the center is a function of the azimuth. On top of the
bar we are advocating here, M31 has two companions (M32 and NGC 205),
both of which can drive spirals and rings. 
 
M32 (NGC 221) is a small companion in the SE quadrant of M31. Baade
believed it to be below the plane of M31 and its systemic velocity is
compatible with
a near-circular orbit which is retrograde with respect to the sense of
rotation of M31. Such an orbit excites an $m$ = 1 inner Lindblad
resonance in the target disc and thus drives a
1-armed leading retrograde spiral, as was shown analytically by
\cite{A78} and confirmed by $N$-body simulations by \cite{Thomason+},
\cite{byrdH} and \cite{Vozik}.
It can also be understood intuitively since a retrograde driving
can not excite any 2-armed spiral\footnote{M32 has a mass of
  only a few percent of that of M31, so that, in order to drive a
  spiral structure 
  whose amplitude is compatible with that observed, it needs to rely on
  a resonance. A simple back-of-the-envelope calculation shows that
  a companion on a retrograde orbit can not induce an $m$=2 resonance,
  only an $m$=1 one.}. So, if
the motion of M32 is indeed retrograde with respect to the 
sense of rotation in M31, one would expect to see a very tightly wound
1-armed leading spiral, comparable to a pseudo-ring, in a region
compatible with the position and 
velocity of M32. Such a spiral component can indeed be found from the
distribution of HII regions and OB associations, as has been shown by
\cite{kalnajs}, \cite{Simien} and \cite{ConsAth}.
This, together with other comparisons with the morphology and
kinematics of M31, show that this is a plausible explanation to the
M31 ring-like structure at a radius of roughly $50'$ \citep{kalnajs,
  athM31conf, Simien}.

\cite{Braun} describes the spatial distribution of the HI gas in M31
by a global two-armed trailing spiral and obtains a good fit if he
allows variations of the orientation of the galaxy plane
both with radius and azimuth (i.e. local line of nodes). He notes that
these mid-plane departures lack bisymmetry and also make the
existence of a massive dark halo unnecessary. He mentions two
possibilities for the driving of the two-armed spiral, namely the
bar-like triaxial bulge \citep[][; etc]{Stark, hoken} and the nearby
companion M32. He favours the second alternative due to the
substantial departures from a planar gas distribution. Since he does
not give a dynamical model providing information on whether the mass
and velocity of M31 are compatible with the observed spiral features,
it is not possible to assess this further. 
 
\cite{Gordon+} also propose a model relying on M32, but without invoking
any resonance. This, however, 
produces only a very large number of arms ($m$ of the order of 10, see
the central right panel of their Fig. 3) and
no ring-like structure, i.e. a structure very unlike what is observed. 
Furthermore, if they use a mass of M32 in 
agreement with observations \citep{mateo}, they find that `very
little effect is seen in the M31 disc', in agreement with what could
be foreseen from the simple explanations given above. They, thus,
artificially increase the mass of M32 by a factor of 5 with respect to
Mateo's estimate, suggesting that M32 could be more massive than
originally thought. 

How does our proposed bar fare with the observed pseudo-ring?
Taking for the semi-major axis length of the bar a value of
$a_b=22'$, as proposed in Sect.~\ref{subsec:denprof}, we can set
constraints 
on the location of corotation, which should necessarily be outside the bar
\citep{contop}. Hydrodynamic simulations \citep{A92b} have
given a range of $(1.2 \pm 0.2)a_b$ for the corotation radius and this
agrees well with 
observational predictions \citep[][etc]{Elmegreen, GKM}.
This places corotation in the range $26' \pm
4'$. Assuming a flat rotation we 
then will have the outer Lindblad resonance of this pattern at $45'
\pm 7'$. It should be noted that this error bar is
based only on the range found from the hydrodynamical simulations,
while, if we we took into account uncertainties on the end of the
bar and on the form of the rotation curve, the error bar would be
considerably larger. Thus, if the rotation curve in the relevant
region is slightly rising \citep[as shown e.g. in][]{cchl},
the location of the outer Lindblad resonance would be somewhat further
out. For example, assuming that the velocity increases roughly as
$r^{0.1}$ places outer Lindblad at $49' \pm 8'$. Barred galaxies are
known to have outer rings and 
both analytical work \citep[e.g.][]{Schwarz} and statistical analysis of the
ring radii \citep*{abcs, buta95} argue that these are
placed at the outer Lindblad resonance. Given the uncertainties, we can
say that the location of the ring in M31 is consistent with its being an
outer ring due to the bar. 

The two alternative explanations leading to a reasonable reproduction
of the M31 ring are thus both
linked to resonances. In the first one, the pseudo-ring is a
very tightly wound one-armed spiral formed by a resonance with M32
rotating in a retrograde orbit around M31. In the second one, it would
be formed at the outer Lindblad resonance of the bar. Is there any way
of discriminating 
between the two alternatives? An argument in favour of the first
explanation is that the observed ring is a pseudo-ring and
its deviations from a ring are more of the $m=1$ than of the $m=2$
type. This, however, is not a strong argument,
since these deviations could be due to the effect of M32
on a pre-existing closed outer ring. An argument in favour of the
second explanation is that it is statistically
more likely, and would make M31 a more ordinary galaxy, its
pseudo-ring in good agreement with a structure
commonly found in a large fraction of barred galaxies. Of course this
is not on its own a sufficient reason to eliminate the first explanation,
particularly since, if M31 was at a larger distance and observed with
lower resolution, the tightly wound one-armed spiral would be confused
with a standard outer ring so that several such structures could be
actually observed and mis-interpreted. A more in depth modeling of
M31 would be necessary to distinguish between the two possible
explanations, or, better still, to produce a model in which both the
effects of the bar and of the companions are taken into account. At
this stage, we can only say that the second 
explanation is statistically more likely, but we still need to model
the effect of M32 on the ring to see whether it is compatible with
the observations. 

\subsection {Relative lengths of the bar and of the boxy bulge}
\label{subsec:bplength}

Because of its viewing angle (near to but not quite
edge-on), M31 brings crucial input to the issue of the relative bar
and boxy bulge lengths. Both orbital structure studies and $N$-body
simulations argue that the extent of the boxy (or equivalently the
peanut, or `X') structure is shorter than that of the bar
\citep*[][A05]{SkokosPatsisAthanassoula2002, PatsisSkokosAthanassoula2002}. 
Observational confirmation, however, is
non-trivial, since the length of the bar can only be measured in 
near face-on galaxies and the length of the boxy bulge only in edge-on
galaxies. 
The main input thus comes from edge-on galaxies on which the extent of
the boxy bulge can be measured directly, while the length of the bar
is inferred from the length of the plateau of the light profile on
a cut along the equatorial plane (e.g. \citeauthor{ldp00b} 
\citeyear{ldp00b}; AM02; \citeauthor{ba05} \citeyear{ba05}; A05). 
This measurement, however, is not 
very accurate, since the end of the plateau tapers off into the
disc. It, nevertheless, provided clear evidence that the bar extends further 
out than the boxy bulge. M31 allows for a different way of comparing
the two extents. Measuring along the major axis, one can delineate
the region where the isophotes are boxy-like and the region where they
have elongations pointing along the bar major axis, i.e. the regions
named in Sect.~\ref{sec:simul} the boxy and the flat bar regions. The
extent of these 
two regions provides us with the length of the boxy bulge and the
length of the bar, respectively. This is well illustrated in
Fig.~\ref{fig:bplength}. Here the length of the bar is obtained from the
face-on views (lower panels) and shown with the full vertical solid
lines, which are continued on the two other panels to show where the end
of the bar lies in the edge-on and M31-like viewing angles. 
The length of the boxy bulge is obtained from the
edge-on view (upper panels) and plotted with the dashed vertical lines,
which are continued on the two other panels to show where the end of
the boxy bulge lies in the face-on and M31-like viewing
angles. Looking at the middle panel we see that the extent of the
regions with the boxy isophotes and the extent of the region with the
elongation towards the bar major axis indeed provide us with the boxy
bulge length and the bar length, respectively. 

Unfortunately, these lengths are not easy to measure directly from the
isophotes. A better determination can be obtained from isophotal fits
and calculation of the $a_4$ and $b_4$ coefficients along them. Such
fits will be made and discussed in Paper II. The measurements we have
here, although not as accurate, are still sufficient for inferring
the type of periodic orbit families which are involved in the building
of M31's box. Using for bar semi-major axis length the value of $22'$
and for the boxy bulge extent the value calculated by \cite{StarkBin},
i.e. $15'$ gives a ratio of 1.4. Although this number
has a large uncertainty, as discussed in the previous sections, it
seems safe to say that it is less than 2. This limit is important
since it gives us indications that the main families building the
peanut are those linked to the higher order vertical resonances
(e.g. families x1v3, or x1v4) and excludes the x1v1 family
\citep*{PatsisSkokosAthanassoula2002}. 
It would be important to repeat this calculation for a number
of galaxies seen at roughly the same orientation as M31 and showing
similar features, to see whether the higher order families are always
the most probable building blocks for boxy bulges. 

\subsection{Miscellanea}
\label{subsec:misc}

The comparisons of simulations to observations argue that M31 has
two types of bulges, a boxy/peanut bulge and a classical bulge. In that
it is not unique. In fact, judging from the fact that we see in
many far-from-edge-on early type
barred galaxies both a classical bulge and a strong bar (indicating the
presence of a boxy/peanut bulge), a large fraction of early type
barred galaxies should have the two types of bulges. In late types,
there should be many cases with both a discy bulge (A05) and a boxy/peanut
bulge, but the fraction will be more difficult to assess. Finally,
in many galaxies all three types of bulges could coexist.    

Our two `best' models presented here are not meant to be an exact fit
to M31. We presented four fiducial models to argue that M31 is barred
and to set constraints on its bar properties. We have thus outlined a
class of models which would give reasonable fits. Surely, by looking
through a large number of models within that class it would have been
possible to find some that gave better fits. This task, however, does
not seem warranted for two reasons in particular. First, the free
parameter space is very large. Indeed, little is known on the mass
distribution of the halo (shape and radial profile) and even less
about its velocity distribution. Even for the visible material, there
is a lot of freedom in selecting $Q(r)$. Furthermore, M31 has
 two companions, which will induce asymmetries which any thorough
modeling should also take 
into account. Thus, a full model for M31 is well beyond the scope of
this paper.

\cite{Sofue+}, using B, V, R and I photometry of the center-most parts
of M31, 
revealed a small inner bar of semi-major axis length about $0.'5$.
Its axial ratio is approximately 3 and it has dark lanes on
its leading edges which are slightly curving in the trailing
sense. This small bar has about the same colour and, therefore the
same population as the bulge stars. Such inner bars are found in
the innermost parts of roughly 30\% to 40\% of bars \citep{LSKP, ErSp}.

The local group contains three large spiral galaxies -- namely our own
Galaxy, M31 and M33 -- and all three were initially considered as
unbarred. de Vaucouleurs, based on the so-called forbidden gas
velocities towards the galactic center, proposed first in 1964 that
our galaxy was barred and this is now considered as a well established
fact \citep[for a review see ][ and references therein]{Dehnen2002b}. The first
arguments that M31 is barred came in 1956 \citep{lindblad}, and we have
presented in this paper a number of strong arguments supporting this.
Finally, images in the red, or NIR of M33 show that the spiral arms in
that galaxy do
not reach the center and that in between them there is indication of a
small bar. It thus looks like the two, or perhaps even the three, largest
spirals in our local group are barred. This is in good agreement with
the results found in the NIR from larger samples of disc
galaxies \citep{EskridgeEtal2000, GrosbolPatsisPompei2004, knapen},
showing that the vast majority of spirals is barred,
although in some cases the bar is small.  

The isophotal shapes of M31 are very characteristic of boxy galaxies, providing
straightforward evidence for the existence of the bar. As previously 
discussed, NGC 7582 and NGC 4442
have a similar orientation. In particular, the isophotes of the
latter \citep{BetGal} have characteristics very similar to those of
M31 (see their Fig. 2). As in M31, we can distinguish there the boxy
and the flat bar region, as  
well as a similar skewness. Since these features are easily identified,  
it would be important to study a number
of galaxies seen at roughly the same orientation as M31 and showing
similar characteristics, in order to get more information on the relative
extents of the peanut and the bar and from that infer which
orbital families constitute the building blocks for boxy
bulges.  

\section{Summary}
\label{sec:summary}

In this paper we argue that M31 is a barred galaxy. Its bar is not as
easily distinguished as it is in galaxies nearer to face-on, but it still 
leaves a number of strong clues, both in the photometry and in the
kinematics.
Our arguments are based on comparisons with $N$-body
simulations. Rather than using only one simulation, which would have
allowed us to point out only the generic
bar properties in M31, we use four models with a range of bar
properties. In this way we can argue that there is a bar, as well as
try to constrain the properties of this bar. Our four models
are called MD, MH, MDB and MHB, respectively, for continuity with
previous works. MD has the weakest of the four bars and MH the
strongest. MDB and MHB have a classical bulge as well, contrary to the
other two models which have only the boxy version.

Comparison between the isodensity curves from our
models, projected in the same orientations as M31, and the M31 isophotes
argues very strongly for the existence of a bar. It also allows us to
say that MH is a very unlikely candidate for the M31 bar, since this
model has a characteristic pinching of the 
isophotes in the central parts, totally absent from the observed
isophotes of M31. The MD model is also a rather unlikely candidate, because its
isodensities do not show the required shapes. On the other hand, the
isodensities of our MDB and MHB models reproduce well the observed
isophotes. Both show two distinct regions with characteristic
isophotal shapes. The inner of
the two regions is dominated in the models by the boxy bulge and so we
call it the boxy region. From the similarity of the isodensities and
the isophotes in this region we can infer that in M31 also it is
dominated by the boxy bulge. Outside this region extends in the models
the flat (vertically thin) bar, which gives to the isodensities characteristic
elongations towards a direction near, but somewhat offset from the
galaxy major axis. We call this region the flat bar region. Again
these isophotal shapes are found in the M31 isophotes 
and we can thus attribute them to a bar. The M31 isodensities show
also a characteristic skewness, which can be well reproduced by our
MDB and MHB models provided the angle between the bar and the galaxy major
axes is within a given range of values.

We also compared radial luminosity profiles from M31 -- made along slits
parallel to the galaxy major axis and offset either to the SE, or to
the NW -- to radial density profiles obtained from the models in the
same way. Here again the bar has left a number of clues. The humps and
asymmetries on the observed luminosity profiles argue clearly for the
existence of a strongly non-axisymmetric component, such as a bar. 
Comparisons with the 
$N$-body results allow us to constrain the models. Model MD can not
reproduce observed luminosity profiles, while model MH
reproduces them badly. It is again models MDB and MHB that give good
fits. These comparisons also give constraints on the value of the
angle between the galaxy and the bar major axes and the range thus
obtained agrees well with that obtained with the help of the
isophotes. Furthermore, these comparisons allow to give a rough
estimate of the length of the bar major axis, which we can estimate to
be of the order of $1300''$. 

The PVDs observed in the HI show also characteristic bar
signatures. In particular there are two branches separated by an empty
space. The inner branch is due to gas in the central region and the
second one to gas outside the main bar region. The empty space is due to the
lack of gas in the main bar region. These PVD diagrams also set
constraints on the bar position angle and length. Both are in good
agreement with the values found from the photometry. 

From all these comparisons we can conclude that there exists very strong
evidence that M31 is barred. In our two best models the bar is neither
too strong, nor too 
weak and its major axis has a position angle of roughly 45$^\circ$,
i.e. there is angle of about 20$^\circ$ between the galaxy and the bar
semimajor axes in the plane of the galaxy. These models have both a
classical and a boxy 
bulge. The length of the latter is smaller than that of the bar and
the most probable backbone for the M31 boxy bulge are periodic
families from families of the x$_1$-tree, bifurcating from the x$_1$
family at a higher vertical resonance, such as the the x1v3 or x1v4. 
A bar such as that of our models could create an outer ring, whose
location would be in agreement with the pseudo-ring observed in M31 at
roughly $50'$. This gives a second, alternative, explanation for the
pseudo-ring. 

\section*{Acknowledgments}

We thank A. Bosma, S. Majewski, M. Bureau, P. Patsis and G. Aronica 
for stimulating discussions on boxy bulges and bars. EA thanks
Jean-Charles Lambert for computing assistance. EA thanks the
INSU/CNRS, the region PACA 
and the University of Aix-Marseille I for funds to develop the
computing facilities used for the simulations discussed in this
paper. RLB was funded under National Science Foundation grant
AST-0307842 and a Space Interferometry 
Mission Key Project Grant, NASA/JPL contract 1228235, awarded to
Steven Majewski (Principal Investigator). She  
thanks the University of Virginia Center for Undergraduate Research and
the Harrison Undergraduate Research Awards program.
This publication makes use of data products from the Two Micron
All Sky Survey, which is a joint project of the University of
Massachusetts and the Infrared Processing and Analysis
Center/California Institute of Technology, funded by the National
Aeronautics and Space Administration and the National Science
Foundation.

\label{lastpage}

\end{document}